\documentclass[fleqn,twoside]{article}
\usepackage{espcrc2}
\usepackage{amssymb}
\usepackage{amsmath}
\usepackage{epsfig}
\usepackage{eucal}

\newcommand{\beq}{\begin{equation}} 
\newcommand{\eeq}{\end{equation}}   
\newcommand{\bea}{\begin{eqnarray}} 
\newcommand{\eea}{\end{eqnarray}}

\title{\vspace{-2cm}
\hfill \parbox{5cm}{\normalsize \tt CERN-TH/2003-276 \\ TTP03-41} \vspace{1.3cm} \\
Perspectives for the radiative return
at meson factories~\thanks{Work supported in part by BMBF under 
grant number 05HT9VKB0,
EC 5th Framework Programme under contract
HPRN-CT-2002-00311 (EURIDICE network),
Polish State Committee for Scientific Research (KBN)
 under contract 2 P03B 017 24, 
 TARI project HPRI-CT-1999-00088,
Generalitat Valenciana under grants CTIDIB/2002/24 and GRUPOS03/013, 
and MCyT under grant FPA-2001-3031.}}

\author{Henryk Czy\.z~\address[KAT]{Institute of Physics, 
University of Silesia, PL-40007 Katowice, Poland}\thanks{E-mail: 
{\tt czyz@us.edu.pl}},
Agnieszka Grzeli{\'n}ska~\addressmark
\thanks{E-mail: {\tt agrzel@us.edu.pl}},
Johann H. K\"uhn~\address[KA]{Institut f\"ur Theoretische Teilchenphysik,
Universit\"at Karlsruhe, D-76128 Karlsruhe, Germany}\thanks{E-mail: 
{\tt johann.kuehn@physik.uni-karlsruhe.de}} and
Germ\'an Rodrigo~\address[CERN]
{Theory Division, CERN, CH-1211 Geneva 23, Switzerland}
\address[V]{Instituto de F\'{\i}sica Corpuscular, 
E-46071 Valencia, Spain}\thanks{E-mail: {\tt german.rodrigo@cern.ch};
Supported by EC 5th Framework Programme under contract HPMF-CT-2000-00989.}
}

\begin{document}

\begin{abstract}
The measurement of the pion form factor and, more
generally, of the cross section for electron--positron annihilation
into hadrons through the radiative return has become an important
task for high luminosity colliders such as the $\Phi$- or
$B$-meson factories. This quantity is crucial for predictions of 
the hadronic contributions to $(g-2)_\mu$, the anomalous magnetic 
moment of the muon, and to the running of the electromagnetic coupling.
But the radiative return opens the possibility of many 
other physical applications. The physics potential of 
this method at high luminosity meson factories is discussed,
the last upgraded version of the event generator PHOKHARA is presented,  
and future developments are highlighted.
\end{abstract}

\maketitle

\section{INTRODUCTION}

Electron--positron annihilation into hadrons is one of the basic
reactions of particle physics, crucial for the understanding
of hadronic interactions. At high energies, around the $Z$ resonance,
the measurement of the inclusive cross section and its interpretation
within perturbative QCD~\cite{CKK,HS} give rise to one of the most
precise and theoretically founded  determinations of the strong coupling
constant $\alpha_s$~\cite{EWG}. Also, measurements in the intermediate 
energy region, between 3 GeV and 11 GeV can be used to determine $\alpha_s$
and at the same time give rise to precise measurements of charm 
and bottom quark masses~\cite{KS}. The low energy region is crucial
for predictions of the hadronic contributions to $a_\mu$, the anomalous 
magnetic moment of the muon, and to the running of the electromagnetic
coupling from its value at low energy up to $M_Z$.

Last, but not least, the investigation of the exclusive final states
at large momenta allows for tests of our theoretical 
understanding of form factors within the framework of perturbative QCD.
Beyond the intrinsic interest in this reaction,
these studies may provide important clues for the interpretation
of exclusive decays of B- and D-mesons, a topic of evident 
importance for the extraction of CKM matrix elements. 

\section{HADRONIC CONTRIBUTIONS TO $a_\mu$ AND $\alpha_\mathrm{QED}$}

The main uncertainty to $a_\mu$ and $\alpha_\mathrm{QED}$ is driven 
by their respective hadronic contributions, which 
are estimated though dispersion integrals 
\begin{eqnarray} 
&& \!\!\!\!\!\!\!\!\!
a_{\mu}^{\mathtt{had},\mathrm{LO}}
= \frac{\alpha^2}{3 \pi^2} \int_{4m_{\pi}^2}^{\infty}  
\frac{ds}{s} \; K(s) \; R(s)~, \\
&& \!\!\!\!\!\!\!\!\!
\Delta \alpha_{\mathtt{had}}(m_Z^2) = -\frac{\alpha m_Z^2}{3 \pi} \ 
\mathrm{Re} \int_{4m_{\pi}^2}^{\infty}  
\frac{ds}{s} \; \frac{ R(s)}{s-m_Z^2}~, \nonumber
\end{eqnarray} 
where the spectral function 
\begin{equation}
R(s)\propto|\langle0|J_\mu|\mathrm{had},(\gamma)\rangle|^2~,
\end{equation} 
is obtained from experimental data of the reaction $e^+e^- \to$ hadrons.
The most recent experimental result for $a_\mu$~\cite{Bennet}
shows a 1.9$\sigma$ discrepancy with respect to the SM prediction 
for this quantity~\cite{Davier:2002dy,HMNT02,Ghozzi:2003yn}. 
Alternatively, one can also use current conservation (CVC)
and isospin symmetry to obtain $R(s)$ from $\tau$ decays.   
In the latter, a 0.7$\sigma$ discrepancy is found~\cite{Davier:2002dy},
which however is incompatible with the $e^+e^-$ based result.
Unaccounted isospin breaking corrections due to the difference of the 
mass and width of the neutral to the charged $\rho$-meson
could explain this discrepancy~\cite{Ghozzi:2003yn}, leaving 
the $e^+e^-$ based analysis as the most reliable.

\section{RADIATIVE RETURN AT MESON FACTORIES}
 
The recent advent of $\Phi$- and $B$-meson factories
allows us to exploit the radiative return to explore the hadronic 
cross section in the whole energy region from threshold up to 
the nominal energy 
of the collider in one homogeneous data sample~\cite{Binner:1999bt,Zerwas}. 
The radiative suppression factor ${\cal O}(\alpha/\pi)$ is easily 
compensated at these factories by their enormous luminosity. 

\begin{figure}[htb]
\begin{center}
\epsfig{file=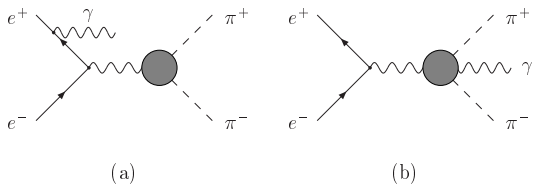,width=7.5cm} \vspace{-.7cm}
\caption{Leading order contributions to the reaction 
$e^+e^-\to\pi^+\pi^-\gamma$ from ISR~(a) and FSR~(b).}
\label{fig1}
\end{center}
\end{figure}

\begin{figure}[htb]
\epsfig{file=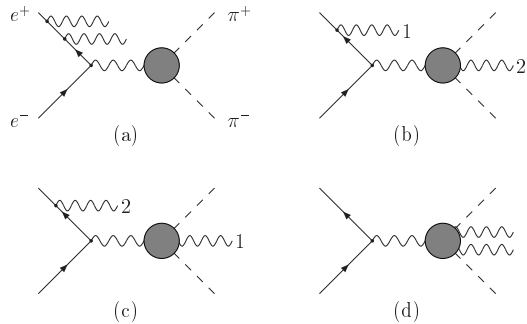,width=7.5cm} \vspace{-.7cm}
\caption{Typical amplitudes contributing to the reaction
$e^+e^-\to\pi^+\pi^-\gamma\gamma$. For two photons emitted either from
the electron/positron or the hadronic system. Only one representative is
displayed.}
\label{fig8}
\end{figure}

\begin{figure}[htb]
\epsfig{file=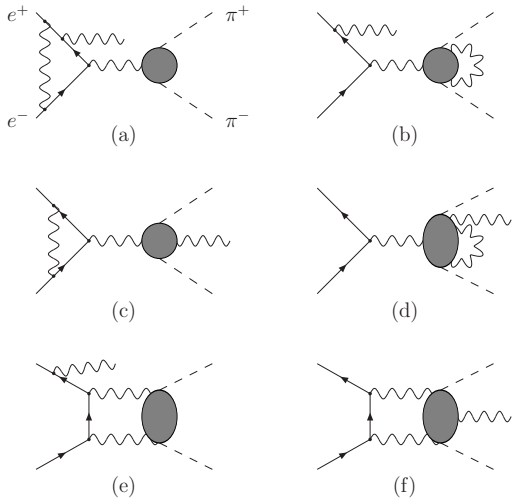,width=7.5cm} \vspace{-.7cm}
\caption{Typical amplitudes describing virtual corrections to the reaction
$e^+e^-\to\pi^+\pi^-\gamma$. Permutations are omitted.}
\label{fig9}
\end{figure}

In principle, the reaction $e^+e^- \to \gamma + {\mathrm {hadrons}}$ 
receives contributions from both initial- and final-state radiation 
(Fig.~\ref{fig1}), ISR and FSR respectively. Only the former is of 
interest for the radiative return.
A variety of methods to disentangle FSR from the ISR contribution 
have been described in detail in~\cite{Binner:1999bt,Rodrigo:2001kf,Czyz:2002np,Czyz:2003ue,Kuhn:2001}.
The first possibility is based on the fact that FSR is dominated by photons
collinear to $\pi^+$ or $\pi^-$, ISR by photons collinear to the beam
direction. This suggests that one should consider events with photons  
well separated from the charged pions and preferentially close to the 
beam. The second option is based on the markedly different angular 
distributions of ISR and FSR, and the characteristic feature of 
their interference. Various charge-asymmetric distributions can be used 
for independent tests of the FSR model amplitude, with the 
forward--backward asymmetry as simplest example. Note however that 
at $B$-factories the $\pi^+\pi^-\gamma$ final state is completely 
dominated by ISR.

Higher order radiative corrections (see Figs.~\ref{fig8} and~\ref{fig9}) 
are also important for the 
precise extraction of the hadronic cross section through the 
radiative return. Furthermore, simulation of individual 
exclusive channels: e.g. $\pi^+\pi^-$, $K^+K^-$, $p\bar{p}$, 
$\pi^0\pi^+\pi^-$, 4$\pi$ , 5$\pi$, 6$\pi$, $\pi\pi\eta$, 
$K\bar{K}\pi$,  $K\bar{K}\pi\pi$, $2K2\bar{K}$, $K\bar{K}\eta$,      
in the low energy region ($<$ 2-3 GeV) requires a fairly detailed 
parametrization of various form factors.

\section{MONTE CARLO EVENT GENERATORS}

The proper analysis requires necessarily the construction of Monte Carlo 
event generators. The event generators EVA~\cite{Binner:1999bt} and 
EVA4$\pi$~\cite{Czyz:2000wh} were based on a leading order treatment of 
ISR and FSR, supplemented by an approximate inclusion of additional 
collinear radiation based on structure functions.
Subsequently, the event generator PHOKHARA was 
developed~\cite{Rodrigo:2001kf,Czyz:2002np,Czyz:2003ue}; it is based
on a complete next-to-leading order (NLO) treatment of 
radiative corrections~\cite{Rodrigo:2001jr,Kuhn:2002xg}.
In its version 2.0 it included ISR at NLO and FSR at LO
for $\pi^+ \pi^-$ and $\mu^+ \mu^-$ final states, and four-pion
final states (without FSR) with some improvements with respect 
to the formulation described in~\cite{Czyz:2000wh}.

The most recent version of PHOKHARA, version 3.0~\cite{Czyz:2003ue}, 
allows for the simultaneous emission of one photon from 
the initial and one photon from the final state. 
This includes in particular the radiative return
to $\pi^+ \pi^- (\gamma)$ and thus the measurement of the (one-photon) 
inclusive $\pi^+ \pi^-$ cross section, an issue closely 
connected to the question of $\pi^+ \pi^- (\gamma)$ 
contributions to $a_{\mu}$.

Recently, a new Monte Carlo event generator, EKHARA~\cite{Czyz:2003gb},  
has been constructed to simulate the reaction $e^+e^- \to \pi^+\pi^-e^+e^-$,
a potential background of the radiative return in particular 
at lower energies.  

Experimental studies presented in \cite{Pisa,unknown:2003jn,Auber,Davier,Simon}
indeed demonstrate the power of the method and seem to indicate 
that a precision of one per cent or better is within reach.
In view of this progress a further improvement of theoretical 
understanding and Monte Carlo generators is required.

\section{PHOTONIC CORRECTIONS TO $a_\mu$}
\label{sec:amu}

\begin{figure}[htb]
\begin{center}
\epsfig{file=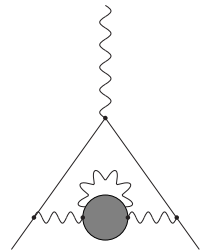,width=4cm} 
\caption{Photonic corrections to $a_\mu$.}
\label{figbomb}
\end{center}
\end{figure}

The issue of photon radiation from the final states is closely 
connected to the question of photonic
contributions to $a_{\mu}$, see Fig.~\ref{figbomb} 
(for related discussions, see e.g.~\cite{Melnikov:2001uw,Gluza:rad}).
At the present level of precision of hadronic contributions to
$a_\mu$, roughly half to one per cent, these corrections 
start to become relevant. 

Qualitatively the order of magnitude of this effect can be
estimated either by using the quark model with $m_u\approx m_d\approx
m_s \approx 180$~MeV (adopted to describe the lowest order 
contribution~\cite{piv}), or with  $m_u\approx m_d\approx m_s \approx 66$~MeV
(adopted to describe the lowest order contribution to $\alpha(M_Z)$) 
or by using $\pi^+\pi^-$ as dominant intermediate hadronic state plus 
photons coupled according to sQED. The three estimates 
(see also~\cite{Melnikov:2001uw})
\bea
\delta a_\mu({\rm quark},\gamma,m_q=180~{\rm MeV})
 = \ 1.880 \times 10^{-10}~, \nonumber && \\
\delta a_\mu({\rm quark},\gamma,m_q=66~{\rm MeV})
 = \ 8.577 \times 10^{-10}~, \nonumber && \\
 \delta a_\mu(\pi^+\pi^-,\gamma) 
 = \ 4.309 \times 10^{-10}~, \nonumber && 
\eea
are comparable in magnitude and begin to be relevant at the present
level of precision of $\pm 8\cdot 10^{-10}$. This order-of-magnitude estimate
suggests that a more careful analysis is desirable.

\begin{figure}[ht]
\epsfig{file=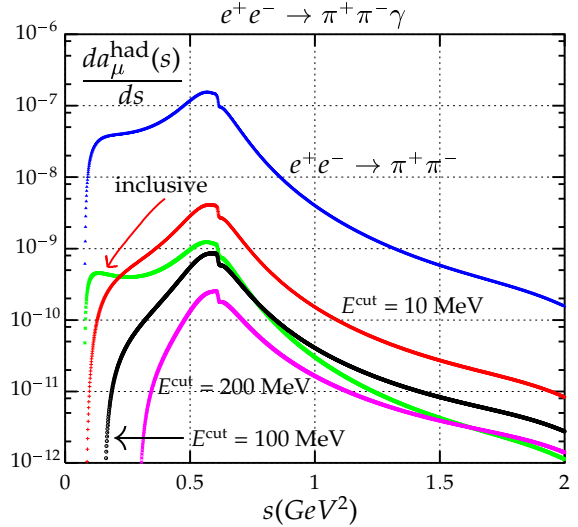,width=7.5cm, height=7cm} \vspace{-1cm}
\caption{Differential contribution to $a_\mu^\mathrm{had,\gamma}$ 
from $\pi^+\pi^-\gamma$ intermediate states for different cutoff values 
compared with the complete contribution (virtual plus real corrections, 
labeled `inclusive') evaluated in sQED (FSR), as well as 
with the contribution from the $\pi^+\pi^-$ intermediate state.}
\label{fig6}
\end{figure}

\begin{figure}[ht]
\epsfig{file=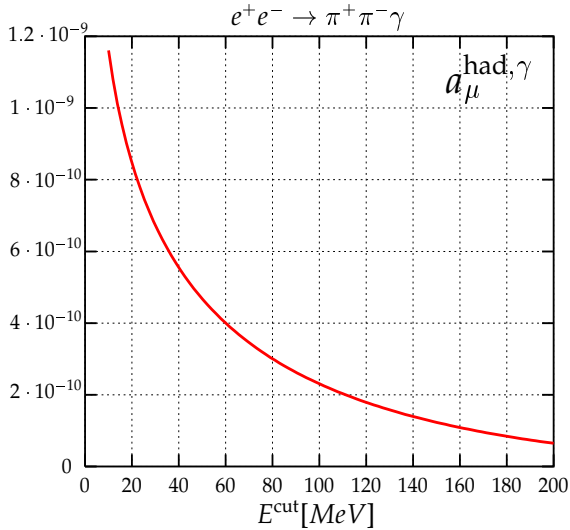,width=7.5cm, height=7cm}
\caption{Integrated contribution to $a_\mu^\mathrm{had,\gamma}$  
as a function of the cutoff $E^\mathrm{cut}$.}
\label{fig7}
\end{figure}

Soft photon emission is clearly described by
the point-like pion model. Hard photon emission, with 
$E_{\gamma} \geq {\cal O}$(100 MeV), however, might be sensitive
to unknown hadronic physics. Therefore the size of
virtual, soft and hard corrections has to be studied separately.
Let us now estimate the contributions from hard photon radiation to
$a_\mu$. In Fig.~\ref{fig6} we display the integrand
\begin{equation}
\frac{d}{ds} a^\mathrm{had,\gamma}_\mu(E^\mathrm{cut}) = 
\frac{\alpha^2}{3\pi^2 s} K(s) \ R^{\mathrm{H}}(s,E^\mathrm{cut})~,
\end{equation}
for $E^\mathrm{cut} =10$, $100$ and $200$~MeV as a function of $\sqrt{s}=
m(\pi^+\pi^-\gamma)$ between the threshold and $2$~GeV. The result is
compared with the complete sQED contribution, as derived from point-like
pions, and the lowest order contribution from $\pi^+\pi^-$. 
The integrated result, with $E_\gamma$ between $E^{{\rm cut}}$
 and infinity,
 is displayed in Fig.~\ref{fig7}.
Contributions from the hard region, above 100 MeV, are clearly small 
with respect to the present experimental and theory-induced uncertainty.

A measurement of $\gamma^*\to \pi^+\pi^-\gamma$ for variable 
$\sqrt{s}$ is desirable for an independent cross check.
This is indeed possible with the radiative return, if events with 
two photons in  the final state are investigated~\cite{Czyz:2003ue}. 
This was in fact one of the motivations for extending the
event generator PHO\-KHA\-RA to events with simultaneous ISR 
and FSR.

\section{PERSPECTIVES FOR THE RADIATIVE RETURN}

Let us now describe a number of topics for the development 
of PHOKHARA \cite{Rodrigo:2001kf,Czyz:2002np,Czyz:2003ue}, 
which are of course intimately connected to the future
potential of measurements through the radiative return.
To large extent this is still a wish list, which will at best be partially
fulfilled, depending also on the future development 
in the experimental analysis.
Some of these topics are based on fairly straightforward extensions of 
the present framework, some would require major calculations.
Inclusion of additional hadronic final states is only possible on 
the basis of  parametrizations (not yet available) for 
the production amplitudes of multi-hadronic states. The radiative
return may, finally, give access to completely new phenomena.
For example one may scrutinize the $D^0\bar{D^0}$-system, produced in 
the radiative return to the $\psi''$, for time-dependent mixing
effects, which are not easily accessible in other experiments.

\subsection{THE RADIATIVE RETURN TO MUON PAIRS \\}

The present version of PHOKHARA (3.0) includes the amplitudes for
\begin{equation}
\kern+60pt e^+e^- \to \mu^+ \mu^- \gamma~,
\end{equation}
in Born approximation plus those ${\cal O}(\alpha)$ real and virtual
corrections which are attached to the electron side 
\cite{Rodrigo:2001kf,Czyz:2002np,Czyz:2003ue}.
Important additional corrections are expected from the second
order process, where one hard photon is emitted from 
the electron, another photon is collinear with the muon.
These additional terms are presently being included in
the simulation \cite{CGGK}.
Resonances, narrow ones like $J/\psi$, wide ones like 
the $\rho$-meson and the vacuum polarization from virtual
hadronic and leptonic intermediate states affects the 
$\mu$-pair cross section. In particular the effects of the vacuum 
polarization should and will be included in a detailed comparison
between theory and experiment.

\subsection{BENCHMARK TESTS AND COMPARISONS \\}

To arrive at reliable Monte Carlo programs suited for the 
experimental analysis comparisons between the results of
simulations, semi-analytical results and comparisons between
programs of different authors are of paramount importance.
Inclusive semianalytical results of ISR NLO corrections 
\cite{Berends:1986yy,Berends:1988ab} are in agreement
with PHOKHARA at a per mill level or better  \cite{Czyz:2002np}.
Preliminary studies \cite{KKMC}
indicate agreement between PHOKHARA~3.0 \cite{Czyz:2003ue} 
and KKMC \cite{KKMC1} for
the $\mu^+\mu^-\gamma$ mode at the level of two per mill in
the region of interest, at least in the mode where FSR had been
switched off.
 The agreement of partial results \cite{Babayaga1} between PHOKHARA
 and BABAYAGA \cite{Babayaga} event generators in $\pi^+\pi^-\gamma$
 modes is encouraging.
Semi inclusive analytical results with specific idealized cuts have 
been published in \cite{Khoze}. However, no comparison with already
available numerical results of \cite{Rodrigo:2001kf} or with the Monte Carlo 
program PHOKHARA has been presented.
Clearly, all that can only be the starting point 
for more detailed studies.

\subsection{HADRONIC FINAL STATES}

\kern+10pt {\bf i) Two mesons:} e.g $K^+K^-$ and $K^0\bar{K^0}$

At present only two-pion and four-pion final states are implemented,
and the parametrizations of the form factors are optimized
for $Q^2$ below 3 to 4~GeV$^2$. Larger values of $Q^2$,
up to ${\cal O}$(10~GeV$^2$) are at reach at B-meson factories.
Although the implementation of amplitudes and form factors
for arbitrary two meson final states is straightforward once
the generator has been coded for $\pi^+\pi^-$, 
the remaining two-body final states $K^+K^-$ and $K^0\bar{K^0}$
are still missing, as well as a reliable and theoretically
well founded formulation for the pion form factor which is
applicable both in the $\rho$ and $\rho'$ resonance region 
and for large $Q^2$ at the same time.

\kern+10pt {\bf ii) Three meson:} e.g. $3\pi$, $\pi\pi\eta$
and $KK\pi$

Strangeness-, isospin- and CP-conservation alone 
lead to significant restrictions
on the structure of production amplitudes for three meson final states.
CP- and G-parity restricts the three-pion state (and all states
with an odd number of pions) to the isospin zero
configuration, dominated by the $\omega$-resonance and its radial
excitations. (The $\phi$ as well as the $J/\psi$ contributions
are suppressed as a consequence of Zweig's rule, but not strictly
forbidden.) The $\pi\pi\eta$ (and $\pi\pi\eta'$) mode is restricted
to the isospin one channel with a production amplitude modeled 
according to the process 
\begin{equation}
\kern+30pt \gamma^* \to \rho \to \eta \rho~ (\to \pi\pi)~.
\end{equation}
Here $\rho$ stands for the full set of $\rho$-meson-like
excitations and the amplitude is written as
\begin{eqnarray}
&\kern-134pt\langle \pi\pi\eta|J_\mu|,0\rangle\propto 
F_{\rho}\left(Q^2\right) F_{\rho}\left((p_{+}+p_{-})^2\right) 
\times \nonumber \\
 \epsilon_{\mu\nu\alpha\beta}
Q^\alpha (p_{+}^\beta+p_{-}^\beta) (p_{+}^\nu-p_{-}^\nu)~.
\end{eqnarray}
The $KK\pi$ mode may receive contributions from isospin zero
and isospin one with a rich structure of $\rho$- and $K^*$-
resonances in the $K\bar{K}$ and $K\pi$ subchannels respectively.
The parametrization of amplitudes for final states with
higher multiplicities, involving $K$, $\pi$, and $\eta$ will
quickly lead to enormously complicated resonance structures and
only data-driven analysis can lead to reliable parametrization of 
the corresponding amplitudes.

\kern+10pt {\bf iii) Baryons:}

The situation is simple for the measurement of the proton
(and neutron) electric and magnetic form factors,
\begin{eqnarray}
 \langle p\bar{p}|J_\mu|,0\rangle = \gamma_{\mu} F_{1}+ 
\sigma_{\mu\nu} \left( p_{+}^{\nu}+p_{-}^{\nu} \right)
\frac{F_2}{4m}~, \nonumber \\ 
&\kern-200pt G_M = F_1 + F_2~,, \nonumber\\
&\kern-200pt G_E = F_1 + \tau F_2~,
\end{eqnarray}
with $\tau  \equiv Q^2/4m_{B}^2>1$.
\begin{figure}[htb]
\vspace{9pt}
\begin{center}
\includegraphics[width=7cm]{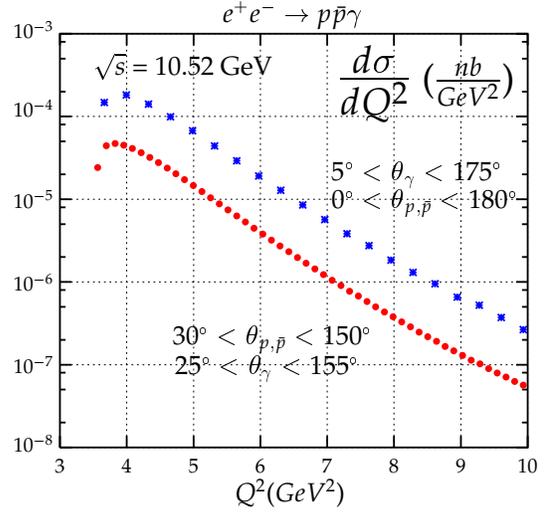} 
\caption{Differential in Q$^2$ cross section of the process
$e^+e^- \to p\bar{p}\gamma$ for two different sets of cuts.}
\label{fig:pp1}
\end{center}
\end{figure}
Sizeable event rates can be reached out to fairly high $Q^2$ 
(see Fig.\ref{fig:pp1}) and
the angular distributions allow for a clean separation of the two
form factors \cite{Proton}. 

\kern+10pt {\bf iv) Multi-particle continuum:} 

Up to now we have restricted the discussion to
exclusive final states. The situation may become even more
complicated for the multi-particle continuum at larger $Q^2$,
say above $\cal O$(10 GeV$^2$). In the parton model this reaction
would be modeled by $e^+e^- \to \gamma q \bar{q}$.
Hadronisation of the $q\bar{q}$ state could be simulated in 
a phenomenological way. However, final state radiation
will be an important background, and elaborate methods,
based e.g. on the markedly different angular distributions of ISR
versus FSR will have to be developed to analyze and separate 
the two contributions [see e.g. \cite{Czyz:2003ue} Eq.(3)]. 
In principle this method
could also be employed to measure the R-ratio above charm 
threshold. Making use of the large boost of 
the D-mesons and their displaced decay vertices, 
the separation of charm and non-charm contributions might
be feasible, thus providing an important input for 
the determination of m$_c$ through sum rules \cite{Kuhn:2002}.
From the purely technical viewpoint PHOKHARA can already now 
be used for the simulation of this reaction, by simply replacing
muon mass and charge by the corresponding values for the charm 
quarks (Note, that a sizeable forward-backward asymmetry is predicted
from ISR-FSR interference also in this case.).

\subsection{NARROW RESONANCES \\}

The production of narrow bottonium states through the radiative return
has been discussed in some detail in \cite{Teubner} 
(see also \cite{Benayoun,cleo}). 
Here we shall concentrate on the $J/\psi$-resonance and its radial
excitations. Neglecting for the moment radiative corrections
and interference between the resonance and the continuum amplitude,
the cross section is given by
\begin{eqnarray}
\sigma_{R} &=& \frac{\Gamma_e}{M}\frac{12\pi\alpha}{s} \biggl\{ 
\frac{1+(M^2/s)^2}{1-(M^2/s)^2}\log\biggl(\frac{1+\cos\theta_m}{1-\cos\theta_m}
\biggr)  \nonumber \\
&-& \biggl(1-\frac{M^2}{s}\biggr) \cos\theta_m \biggr\}~. 
\label{sigma}
\end{eqnarray}

The angular region of the photon is restricted by 
$|\cos\theta|\leq\cos\theta_{m}$ and $\Gamma_e$ denotes
the resonance  partial decay rate into $e^+e^-$. For an integrated
luminosity of 100fb$^{-1}$ this corresponds to 34$\cdot$10$^4$
 $J/\psi$ events,
out of which 20000 events will be found in the $\mu^+\mu^-$ channel.
It is remarkable, that the results form the
BABAR-collaboration for this reaction (see \cite{Auber,Davier})
have lead to a value for $\Gamma_e \Gamma_{\mu}/\Gamma_{tot}$ 
which already now is more precise than the world average as
compiled in \cite{PDG}. The enormous statistics gives access to 
a multitude of final states. In particular it is possible to measure
simultaneously the cross section on- and off-resonance.
For the double ratio
\begin{equation}
 RD_f \equiv \frac{\sigma_f}{\sigma_{\mu+\mu-}}(\mathrm{on~res.})
/ \frac{\sigma_f}{\sigma_{\mu+\mu-}}(\mathrm{off~res.})~,
\end{equation}
one evidently expects RD$_f$ = 1, if the decay $J/\psi \to f$  can
proceed through the virtual photon only. This is expected
for final states with isospin 1, like $2\pi$, $4\pi$, $6\pi$
or $\pi\pi\eta$. Final states with isospin zero, like $3\pi$, 
$5\pi$, $K\bar{K}$, $K\bar{K}\pi$ are dominated by hadronic $J/\psi$
decays and, in the framework of perturbative QCD mediated by
the three gluon intermediate state.

\subsection{RADIATIVE PRODUCTION \\  OF $\psi(3770)$ \\}

A wealth of physics results has bee obtained from $B\bar{B}$ 
production at asymmetric $B$ factories. The $B\bar{B}$ state is
produced in a 1$^{--}$ configuration, which leads to nontrivial
correlations between the decay products of $B$ and $\bar{B}$
mesons. The Lorentz boost allows to resolve the correlated
time dependence of the two decays, which in turn is the basis for
the measurement of the CP asymmetry. Similar studies can, 
in principle be performed for the $D\bar{D}$ system with the help
of the radiative production of $\psi$(3770) and its subsequent
decay into $D\bar{D}$. From Eq.(\ref{sigma}), using 
$\Gamma_e = $ 0.26 keV
and assuming an integrated luminosity of 100~fb$^{-1}$ one expects
a total sample of 15000 $\psi$(3770) events. 
Assuming an integrated luminosity of order 10~ab$^{-1}$, 
as suggested for future upgrades
of $B$ factories, more than a million of $\psi$(3770) could be produced.
Let us discuss a number of potentially interesting studies.
The relative amount of charged versus neutral final states
in the decays of $\phi \to K\bar{K}$, $\psi(3770)\to D\bar{D}$
and $\Upsilon(4S) \to B\bar{B}$ and its energy dependence
around the resonance is sensitive to phase space effects,
Coulomb attraction and hadronic final state 
interactions. The measurement of $\psi$(3770) in the radiative return
and its separation into charged and neutral modes automatically
involves the full energy range.

Let us now speculate about the potential of such
a measurement, once millions of $\psi$(3770) are available \cite{Mannel}.
Tagging the flavour of one $D$ meson through the leptonic decay
of the other meson, all those studies are feasible, 
which can also be investigated in the decay of neutral $D$ mesons,
which are tagged through the charge of the pion in the
$D^* \to \pi D$ decay. Mixing would manifest itself through like
sign dilepton events from $D \bar{D}$ decays, and the 
time dependence of this phenomenom could be studied in 
the present setup. Another option would be the study of 
the time dependence of doubly Cabbibo suppressed decays.
A different subject, specific to $\psi$(3770) decay, is 
the investigation of CP eigenstates in the decay of both
$D^0$ and $\bar{D}^0$ and the corresponding time evolution.
Assuming CP-conservation 
the combination $D^0(\to \pi^+\pi^-)\bar{D}^0(\to \pi^+\pi^-)$ is
for example strictly forbidden, independent of the question
of $D \bar{D}$ mixing. In contrast to $\psi$(3770) production 
at a symmetric charm factory it would be possible to study
not only the rate for such a process but also its time dependence,
which will carry additional information on direct and indirect
CP-violation. Last not least one may use lepton tags on one side
and study the time dependence of decays into CP eigenstates
at the other side, thus exploring combinations quite similar
to those investigated in the $B \bar{B}$ system at asymmetric
$B$-meson factories. Predictions for $D\bar{D}$-mixing and CP
violation within the Standard Model \cite{Bigi} lead to fairly small
effects
which would be difficult to observe. Nevertheless these 
measurements could lead to important limits and perhaps give
access to physics beyond the Standard Model.

\section{CONCLUSIONS}

Measurements of the pion form factor through the radiative return
offer the unique possibility for improved predictions for the
muon magnetic moment and the electromagnetic coupling $\alpha(M_Z)$
but also gives access to many new phenomena.
The Monte Carlo event generator PHOKHARA is being developed 
further to include new features.


\section*{ACKNOWLEDGEMENTS}

It is a pleasure to thank S.~di Falco, M.~Incagli and G.~Venanzoni 
for many discussions on experimental aspects, 
and for the organization and stimulating atmosphere created during 
the workshop.  

We would like to thank A.~Denig, W.~Kluge,
and S.~M\"uller  for discussion of experimental aspects of our
 analysis.

H. Czy\.z and A. Grzeli{\'n}ska
 are grateful for the support and the kind hospitality of
the Institut f{\"u}r Theoretische Teilchenphysik
 of the Universit\"at Karlsruhe.


\end{document}